\documentclass{emulateapj}

\newcommand{\xmm}{{\it XMM-Newton}}
\newcommand{\chan}{{\it Chandra}}
\newcommand{\psr}{PSR~B0355+54}

\shorttitle{The Pulsar Wind Nebula of PSR~B0355+54}
\shortauthors{McGowan et al.}

\begin{document}

\title{Probing the Pulsar Wind Nebula of PSR~B0355+54}

\author{K.E. McGowan\altaffilmark{1,2}, W.T. Vestrand\altaffilmark{3}, 
J.A. Kennea\altaffilmark{4}, S. Zane\altaffilmark{2}, 
M. Cropper\altaffilmark{2}, F.A. C\'ordova\altaffilmark{5}}
\altaffiltext{1} {School of Physics and Astronomy, Southampton University, 
Southampton, UK}
\altaffiltext{2} {Mullard Space Science Laboratory, University College of
London, UK}
\altaffiltext{3} {Los Alamos National Laboratory, Los Alamos, NM 87545}
\altaffiltext{4} {Pennsylvania State University, 525 Davey Laboratory, 
University Park, PA 16802, USA}
\altaffiltext{5} {Chancellor's Office, University of California, Riverside, 
CA 92521}
\email{kem@astro.soton.ac.uk}

\begin{abstract}

We present \xmm\ and \chan\ X-ray observations of the middle-aged radio
pulsar \psr.  Our X-ray observations reveal emission not only from the 
pulsar itself, but also from a compact diffuse component extending 
$\sim 50\arcsec$ in the opposite direction to the pulsar's proper motion.
There is also evidence for the presence of fainter diffuse emission extending
$\sim5\arcmin$ from the point source.  The compact diffuse feature is 
well-fitted with a power-law, the index of which is consistent with the 
values found for other pulsar wind nebulae.  The morphology of the diffuse
component is similar to the ram-pressure confined pulsar wind nebulae detected
for other sources.  The X-ray emission from the pulsar itself is described
well by a thermal plus power-law fit, with the thermal emission most likely
originating in a hot polar cap.

\end{abstract}


\section{Introduction} \label{intro}

Isolated pulsars constitute one of the most powerful laboratories for studying 
particle acceleration in astrophysics.  A significant fraction of the energy 
from rotation-powered pulsars is converted into a wind \citep{ree74}, which 
travels at a velocity close to the speed of light.  The interaction of this 
pulsar wind with the ambient medium produces a shock and acceleration of the 
relativistic particles at the shock generates synchrotron emission.  This 
non-thermal diffuse emission manifests itself as a pulsar wind nebulae (PWN) 
or {\it plerion} at radio and X-ray energies \citep[e.g.][]{ree74,gae01}.
Due to the short synchrotron lifetimes of high energy electrons, X-ray emission
from a PWN directly traces the current energetics of the pulsar.  The spectral
and morphological characteristics of an X-ray PWN therefore reveal the
structure and composition of the pulsar wind and the orientation of the
pulsar's spin axis and/or velocity vector.

The middle-aged 156 ms radio pulsar \psr\ is known to emit X-rays 
\citep{hel83,sew88,sla94} and gamma-rays \citep{bha90}.  \citet{hel83} 
reported the first detection in X-rays of the source using data from 
{\it Einstein}, stating that emission extended $5\arcmin$ from the pulsar.  
However, \citet{sew88} analyzed the {\it Einstein} data and concluded that 
while there was evidence for weak emission $1.7\arcmin$ from the source 
position, emission from the pulsar itself was not detected.  Nevertheless, 
they did not rule out the possibility that the emission could be associated 
with a PWN.  \citet{sla94} detected \psr\ in a 20 ks {\it ROSAT} observation, 
but owing to the lack of counts it was not feasible to perform a spectral 
analysis.  The analysis of the {\it ROSAT} data also led to the detection of 
faint extended emission $\sim 1.6\arcmin$ from the pulsar position, but 
\citet{sla94} did not believe there was enough evidence to support a link 
between the source and the extended emission.

In this paper, we report on \xmm\ and \chan\ observations of \psr\ which we 
use to investigate the presence of diffuse emission that can be attributed to 
a PWN.

\section{Observations and Data Reduction}
\label{obser}

\psr\ was observed with \xmm\ on 2002 February 10 for 29 ks.  We used data 
from the European Photon Imaging Camera (EPIC) PN instrument \citep{str01} for 
the spatial, spectral and timing analysis.  The PN was configured in 
{\it small window} mode and the thin blocking filter was used.  Data from the 
MOS1 instrument \citep{tur01} was also used for the spatial analysis.  The 
MOS1 was operated in {\it full window} mode with the medium filter.  We 
reduced the EPIC data with the \xmm\ Science Analysis System (SAS version 
6.1.0).  In order to maximize the signal-to-noise ratio for our \xmm\ 
observation, we filtered the data to include only single, double, triple and
quadruple photon events for the MOS1, and only single and double photon 
events for the PN.  Data were filtered to exclude events that may be 
incorrect, for example those next to the edges of the CCDs and next to bad 
pixels.  We only included photons with energies in the range $0.3-10$ keV.

\psr\ was observed for 66 ks on 2004 July 16 with the ACIS-S array on \chan\ 
in the very faint timed exposure imaging mode.  We performed standard data 
processing using CIAO version 3.2.  The data were filtered to restrict the
energy range to $0.3-10$ keV and to exclude times of high background.  

\section{Spatial Analysis}
\label{space}

\begin{figure}
\plotone{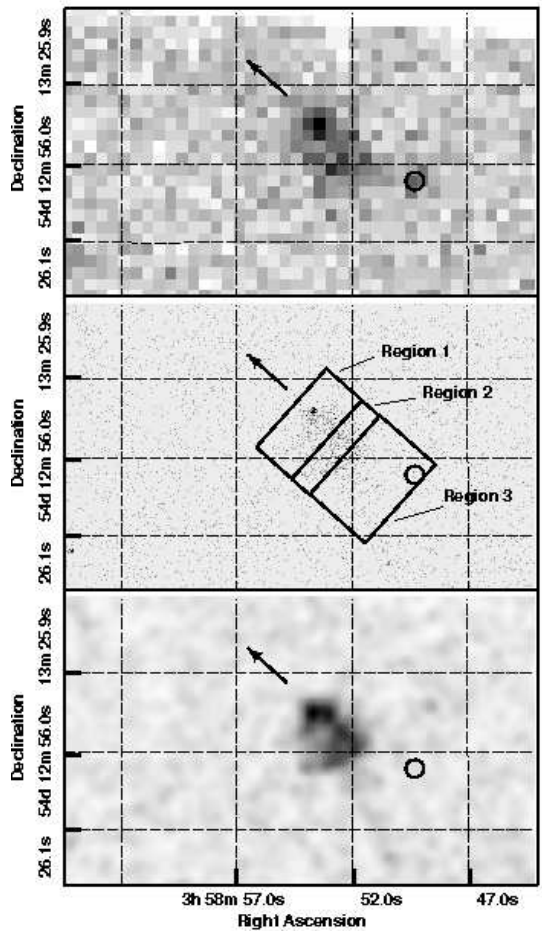}
\caption{X-ray detection of \psr\ and its diffuse emission.  Top panel: 
Gray-scale plot of the $0.3-10$ keV \xmm\ PN image.  Middle panel: Gray-scale 
plot of the $0.3-10$ keV \chan\ ACIS image.  The boxes define the regions 
used to extract spectra for the diffuse emission.  Bottom panel: The same 
image as the middle panel with the contribution from the X-ray point source 
removed.  The image is smoothed with a Gaussian of width $\sim 2\arcsec$.  The 
arrow shows the direction of the proper motion of the pulsar and has a length 
of $20\arcsec$.  The circle marks the ``south west'' source (see Section 3).}
\label{fig_im}
\end{figure}

Initial inspection of the images created from the EPIC-PN and EPIC-MOS1 data 
show relatively strong emission at the pulsar position and evidence for 
extended emission near to \psr\ (see Figure \ref{fig_im}, top panel).  We 
generated a mosaic of the PN and MOS1 images and measured the X-ray source
positions using the SAS source detection tool EDETECT\_CHAIN.  We compared
the positions of the field stars in our observation with the positions
from optical catalogs to determine an astrometric correction.  This correction 
was applied to the X-ray coordinates of the pulsar, resulting in 
R.~A. = $03^{\rm h} 58^{\rm m} 53\fs73$, decl. = $+54\degr 13\arcmin 
12\farcs12$ (J2000), with an rms error of $0\farcs78$.  This position lies
$1\farcs6$ from the radio position.

\begin{figure}
\plotone{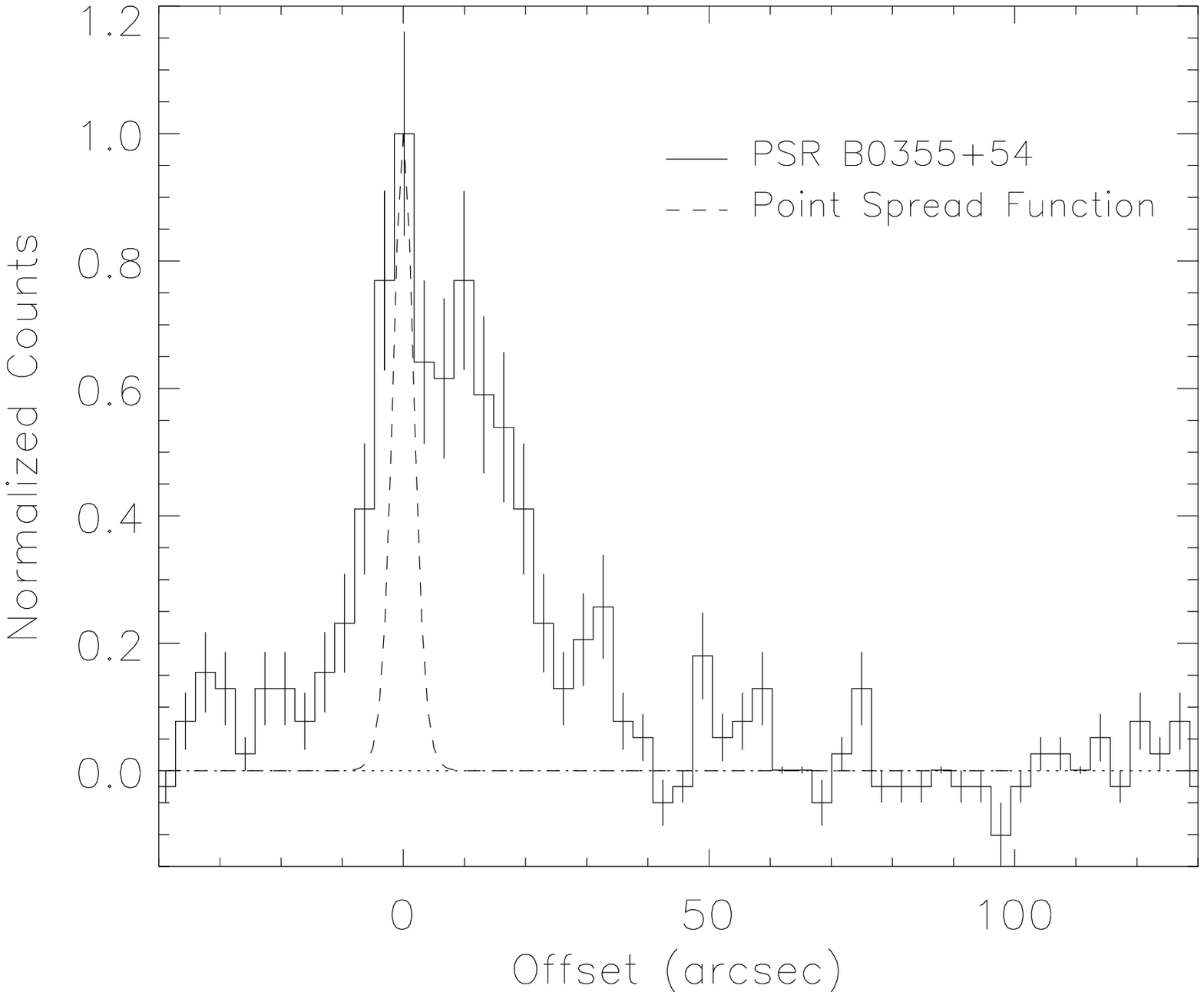}
\plotone{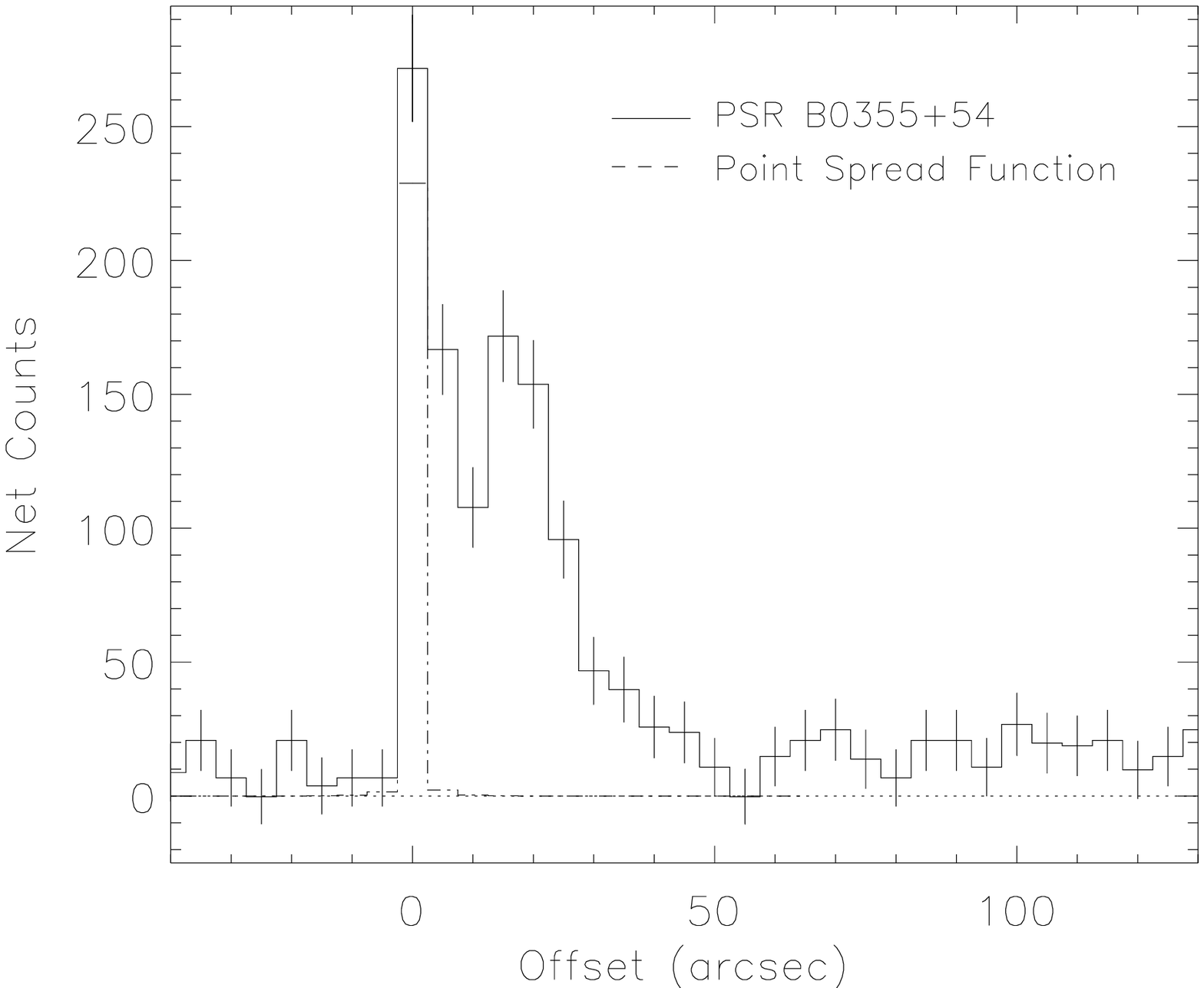}
\caption{X-ray emission from \psr\ as a function of distance from the point 
source along the direction of the proper motion of the pulsar (solid line), 
compared to the instrument PSF determined at 1.5 keV and the location of the 
pulsar (dashed line).  Top panel: \xmm\ PN.  Bottom panel: \chan\ ACIS.}
\label{fig_psf}
\end{figure}

To confirm the presence and examine the extent of the diffuse emission in the 
\xmm\ data we have compared the detected PN emission with that for a point 
source.  We calculated the intensity for the pulsar by using bilinear 
interpolation at regularly spaced points along the direction of proper motion 
of \psr\ \citep{chat04}.  We compared this profile with the \xmm\ point-spread 
function (PSF) for the PN at 1.5 keV, which we generated using the King 
profile parameters included in the \xmm\ calibration file 
''XRT3\_XPSF\_0006.CCF.plt''\footnote{See http://xmm.vilspa.esa.es/docs/documents/CAL-SRN-0100-0-0.ps.gz for more information}.  In Figure \ref{fig_psf} 
(top panel) we show the profiles for the pulsar and the PN PSF.  

The positions of the X-ray sources for the ACIS data were determined using
the CIAO source detection tool WAVDETECT.  The image does not contain enough 
sources with known counterparts to perform an astrometric correction to the 
coordinates.  A point source is detected at R.~A. = $03^{\rm h} 58^{\rm m} 
53\fs70$, decl. = $+54\degr 13\arcmin 13\farcs87$ (J2000), which is 
$0\farcs14$ away from the radio pulsar position.  This source is consistent 
with being the X-ray counterpart of the pulsar.

The ACIS image also reveals a faint tail of emission in the opposite direction 
to the pulsar's proper motion (see Figure \ref{fig_im}, middle and bottom 
panels).  Again we determined the net counts from the source and diffuse 
emission at regularly spaced intervals along the direction of the pulsar's 
proper motion.  We generated a PSF for \psr\ using the \chan\ PSF library 
evaluated at 1.5 keV and the location relevant to our source.  The PSF was 
normalized to the total counts in \psr.  We calculated the net counts for the 
PSF in the same intervals as for \psr.  The source and PSF profiles are shown 
in Figure \ref{fig_psf} (bottom panel).

The \xmm\ PN image shows emission $\sim 45\arcsec$ south west of the pulsar 
which could be a source, however, there is no corresponding emission at this 
position in the \chan\ ACIS image (see Figure \ref{fig_im}).  We performed 
wavelet analysis on the \chan\ data which confirms there is no source 
detected.  Comparison of the \xmm\ field of \psr\ with the Digitized Sky 
Survey does not show any optical source at the position of the south west 
emission.  If the emission in the \xmm\ data is real, it suggests that the 
diffuse emission could be varying over time.  

We smoothed the \chan\ ACIS image with a Gaussian of width $\sim 2\arcsec$ 
(Figure \ref{fig_im}, bottom panel), the resulting image suggests that there 
are two regions of enhanced diffuse emission -- one near to the pulsar and 
the other $\sim 10\arcsec$ away.  The intensity profiles for the \xmm\ and 
\chan\ data indicate that the core of the X-ray emission lies within 
$5 \arcsec$ of the pulsar position.  Both emission profiles indicate that the 
diffuse emission extends out to $\sim 50 \arcsec$, with the bulk of the 
emission lying within $20-30 \arcsec$ of \psr.  The profiles also show 
evidence for a dip in the emission at $\sim 10 \arcsec$ agreeing with the 
smoothed ACIS image.

\begin{figure}
\plotone{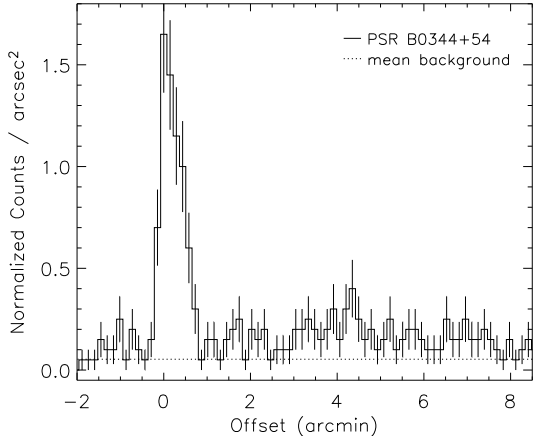}
\caption{\xmm\ MOS1 X-ray emission from \psr\ as a function of distance from 
the point source, orientated $20\degr$ further North from the direction of
the pulsar's proper motion (solid line), compared to the mean background 
(dotted line).}
\label{fig_mos}
\end{figure}

Detection of emission from \psr\ at distances of $1\farcm6$ to $5\arcmin$ 
from the source position i.e. on a larger scale than shown in Figure 
\ref{fig_im}, have been reported by \citet{hel83}, \citet{sew88} 
and \citet{sla94}, see also \citet{tep05}.  Visual inspection of a smoothed 
version of the MOS1 image indicates that there is a region of enhanced 
emission extending a few arcminutes south of the pulsar, orientated
$\sim 20\degr$ closer to North than the diffuse emission we have reported 
above.  The intensity of the MOS1 emission in this direction was determined
using bilinear interpolation at regularly spaced intervals from the source.
We show in Figure \ref{fig_mos} the distribution of counts as a function of 
distance from the source compared to the mean background.  Our analysis 
suggests that there is an excess of counts at $1\arcmin$ and $\sim3-5\arcmin$
from the point source.

\begin{deluxetable*}{lccccc}
\tablewidth{0pt}
\tablecaption{Best-fit parameters for the \xmm\ spectrum of PSR B0355+54 
\label{tab:spec_xmm}} \tablehead{
\colhead{Model} & \colhead{$N_{H}$}  & \colhead{$\Gamma$} &
\colhead{$T/T_{eff}^\infty$} & \colhead{$\chi^{2}_{\nu}$ (dof)}
&\colhead{$F_X$}\\
\colhead{} & \colhead{($10^{22}$ cm$^{-2}$)}  & \colhead{} & \colhead{($10^{6}$ K)}
& \colhead{} & \colhead{(erg cm$^{-2}$ s$^{-1}$)} }
\startdata
PL & $0.50^{+0.36}_{-0.20}$ & $1.5^{+0.5}_{-0.3}$ & \nodata & 0.7 (16) 
&  $(2.3^{+1.0}_{-0.7}) \times 10^{-13}$ \\
BB & $<0.01$ & \nodata & $11.26^{+1.74}_{-1.51}$ & 1.0 (16) & 
$(1.4^{+0.1}_{-1.1}) \times 10^{-13}$ \\
BB + PL & $0.23^{+0.69}_{-0.23}$ & $0.8^{+0.3}_{-0.3}$ & 
$7.66^{+2313}_{-4.64}$ & 0.7 (14) & $(2.0^{+1.0}_{-1.1}) \times 10^{-13}$ \\
NSA & $0.13^{+0.23}_{-0.11}$ & \nodata & $7.66^{+0.01}_{-2.83}$ & 0.9 (16)
& $(1.4^{+0.1}_{-1.3}) \times 10^{-13}$ \\
NSA + PL & $0.49^{+0.21}_{-0.21}$ & $1.5^{+1.0 }_{-0.3}$ & 
$0.64^{+6.83}_{-0.75}$ & 0.8 (14) & $(2.4^{+1.1}_{-1.0}) \times 10^{-13}$ \\
\enddata
\tablecomments{The last column is the unabsorbed flux in the 0.3--10 keV 
range.  In the case of the NSA model the mass and radius of the neutron star 
are fixed at $M_{NS} = 1.4 M_{\sun}$ and $R_{NS} = 10$ km, respectively, and 
the magnetic field of the neutron star is fixed at $B=10^{12}$ G.  
The errors quoted are the 90\% uncertainties.}
\end{deluxetable*}

\section{Spectral Analysis}
\label{spec}

In order to investigate the properties of the X-ray emission from \psr\ and 
the compact ($\leq50\arcsec$) diffuse nebula, we compared the spectra 
extracted from different spatial regions.  Our results from the spatial 
analysis suggest that the core of the pulsar emission lies within $5 \arcsec$ 
of the pulsar's position.  However, in the case of the \xmm\ data, this size 
of aperture does not contain enough counts for a meaningful analysis.  

The pulsar spectrum has been extracted from the \xmm\ observation using a 
circular region of radius $30\arcsec$, centered on the pulsar's radio 
position.  The background was extracted from a region of similar size offset 
from the pulsar position.  The total counts contained in the source region
is 1143 with an estimated 562 from background.  The spectrum was regrouped by 
requiring at least 50 counts per spectral bin.  We created a photon 
redistribution matrix (RMF) and ancillary region file (ARF) for the spectrum.  
The subsequent spectral fitting and analysis was performed using XSPEC, 
version 11.3.1.

\begin{figure}
\plotone{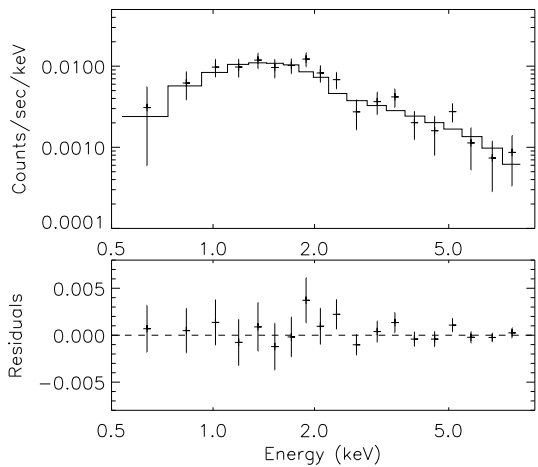}
\caption{The \xmm\ spectrum of \psr\ with best-fit power-law model.  Also 
shown are the residuals from comparison of the data to the model.}
\label{fig:xmm_mod}
\end{figure}

We modeled the spectrum in the $0.5-9.0$ keV range.  Initially we fitted the 
spectrum with single-component models including absorbed power-law, blackbody
and magnetized, pure H atmospheric \citep{pav95} models.  The spectrum is 
best-fitted with a power-law with index $\Gamma = 1.5^{+0.5}_{-0.3}$ and 
column density $N_{H} = (0.50^{+0.36}_{-0.20}) \times 10^{22}$ cm$^{-2}$.  
This value for the power-law index is similar to the values found for other
PWNe \citep[e.g.][]{kas05}.  The Galactic hydrogen column in the direction of 
\psr\ is $N_{H} = 0.88 \times 10^{22}$ cm$^{-2}$.  The fit results in an
unabsorbed 0.3--10 keV energy flux of $(2.3^{+1.0}_{-0.7}) \times 10^{-13}$
ergs cm$^{-2}$ s$^{-1}$.  We also fitted the spectrum with blackbody plus 
power-law and atmospheric plus power-law models, both modified by 
photoelectric absorption.  The multi-component models give similar values for 
reduced $\chi ^2$, however the temperatures implied by the fits are poorly 
constrained.  It is likely that the presence of the pulsar wind nebula, and 
being unable to separate the pulsar core and diffuse emission, effects our 
ability to constrain the thermal component in the spectral fits of the \xmm\ 
data.  The results of the \xmm\ spectral fitting are given in 
Table~\ref{tab:spec_xmm}.  The \xmm\ spectrum with the best-fitting power-law 
model is shown in Figure~\ref{fig:xmm_mod}.

For the \chan\ data we extracted a spectrum for the core of the pulsar 
emission from a circular region of radius $5\arcsec$ centered on the radio 
position.  An annulus centered on the pulsar position was used to extract the 
background, with inner and outer radii of $6\arcsec$ and $10\arcsec$, 
respectively.  We find a total of 244 counts contained in the source 
region, with 29 counts attributed to background.  We created the RMF and ARF 
files using standard CIAO tools.  Before fitting the spectrum we regrouped the
data, requiring a minimum of 15 counts per spectral bin.

We fitted the spectrum in the energy range $0.5-7.0$ keV using the same models
as for the \xmm\ data.  In the first instance we let the neutral hydrogen 
column density be a free parameter; however this led to unreasonably small 
values for $N_{H}$.  Subsequently we fixed the column density at the value 
found from the power-law fit to the \xmm\ spectrum of \psr.  We find that the 
\chan\ spectrum can also be characterized by a power-law.  The model has a 
power-law index of $\Gamma=1.9^{+0.4}_{-0.3}$ which is consistent within the 
90\% uncertainties to the value found from the \xmm\ data.  However, in the 
case of the \chan\ data we find that a thermal plus power-law model provides a 
better fit statistically.  The data are equally well-fitted by a blackbody 
plus power-law and a magnetized, pure H atmospheric \citep[][``nsa'' model in 
XSPEC]{pav95} plus power-law model.  The results of the \chan\ spectral 
fitting are given in Table~\ref{tab:spec_chan}.

\begin{deluxetable*}{llccccc}
\tablewidth{0pt}
\tablecaption{Best-fit parameters for the \chan\ X-ray emission of 
PSR B0355+54 
\label{tab:spec_chan}} \tablehead{
\colhead{Region} & \colhead{Model} & \colhead{$\Gamma$} &
\colhead{$T/T_{eff}^\infty$} & \colhead{$R_{BB}/R_{NS}$} & 
\colhead{$\chi^{2}_{\nu}$ (dof)} & \colhead{$F_X$}\\
\colhead{} & \colhead{} & \colhead{} & 
\colhead{($10^{6}$ K)} & \colhead{km} & \colhead{} & 
\colhead{(erg cm$^{-2}$ s$^{-1}$)} }
\startdata
Core & PL & $1.9^{+0.4}_{-0.3}$ & \nodata & \nodata & 0.5 (34) & 
$(4.9^{+1.6}_{-0.7}) \times 10^{-14}$ \\
 & BB & \nodata & $6.73^{+2.32}_{-1.87}$ & $0.01^{+0.03}_{-0.01}$  & 1.4 (34) 
& $(2.4^{+0.1}_{-0.1}) \times 10^{-14}$ \\
 & BB + PL & $1.0^{+0.2}_{-1.0}$ & $2.32^{+1.16}_{-0.81}$ & 
$0.12^{+0.16}_{-0.07}$  & 0.3 (32) & $(6.4^{+19.3}_{-1.1}) \times 10^{-14}$ \\
 & NSA & \nodata & 0.48 & 9.5 & 3.6 (34) & 
$(1.8^{+0.3}_{-0.5}) \times 10^{-13}$ \\
 & NSA + PL & $1.5^{+0.5}_{-0.4}$ & $0.45^{+0.20}_{-0.22}$ & 
$7.2^{+7.2}_{-2.2}$ & 0.4 (32) & $(1.5^{+54.0}_{-0.7}) \times 10^{-13}$ \\
Diffuse -- all & PL & $1.4^{+0.3}_{-0.3}$ & \nodata & \nodata & 1.0 (50) &
$(1.7^{+0.8}_{-0.5}) \times 10^{-13}$ \\ 
Diffuse -- 1 & PL & $1.4^{+0.4}_{-0.4}$ & \nodata & \nodata & 1.0 (16) &
$(5.7^{+3.3}_{-1.8}) \times 10^{-14}$ \\
Diffuse -- 2 & PL & $1.5^{+0.3}_{-0.3}$ & \nodata & \nodata & 1.4 (16) &
$(7.6^{+3.4}_{-1.9}) \times 10^{-14}$ \\
Diffuse -- 3 & PL & $1.2^{+0.5}_{-0.4}$ & \nodata & \nodata & 1.2 (17) &
$(5.3^{+4.4}_{-2.0}) \times 10^{-14}$ \\
\enddata
\tablecomments{The neutral hydrogen column density has been fixed at 
$N_{H}=0.50 \times 10^{22}$ cm$^{-2}$ in all of the fits.  The last column is 
the unabsorbed flux in the 0.3--10 keV range.  In the case of the NSA model 
the distance to the source is fixed at $D=1.04$ kpc, the mass of the neutron 
star is fixed at $M_{NS} = 1.4 M_{\sun}$, and the magnetic field of the 
neutron star is fixed at $B=10^{12}$ G.  The errors quoted are the 90\% 
uncertainties.}
\end{deluxetable*}

\begin{figure}
\plotone{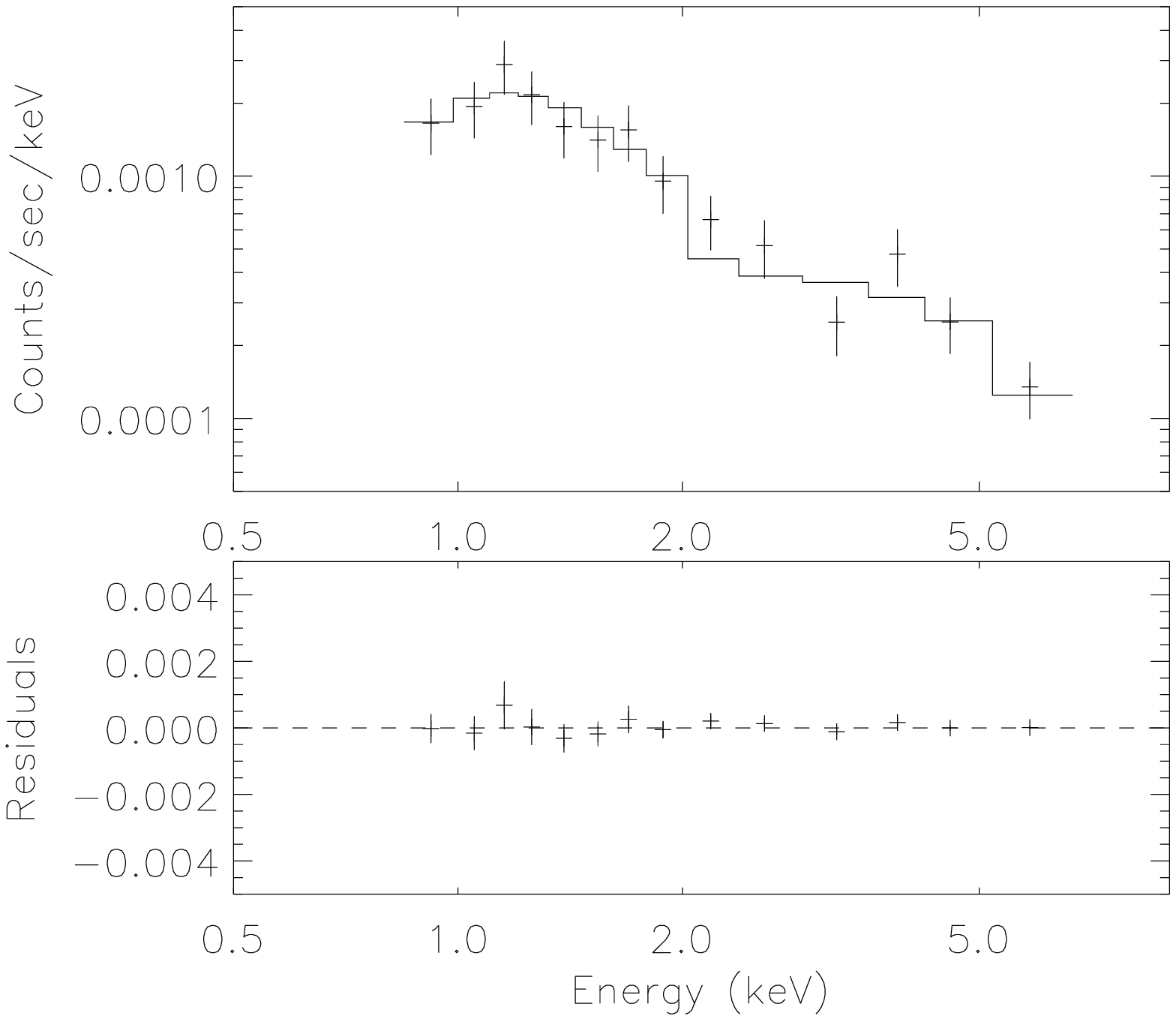}
\plotone{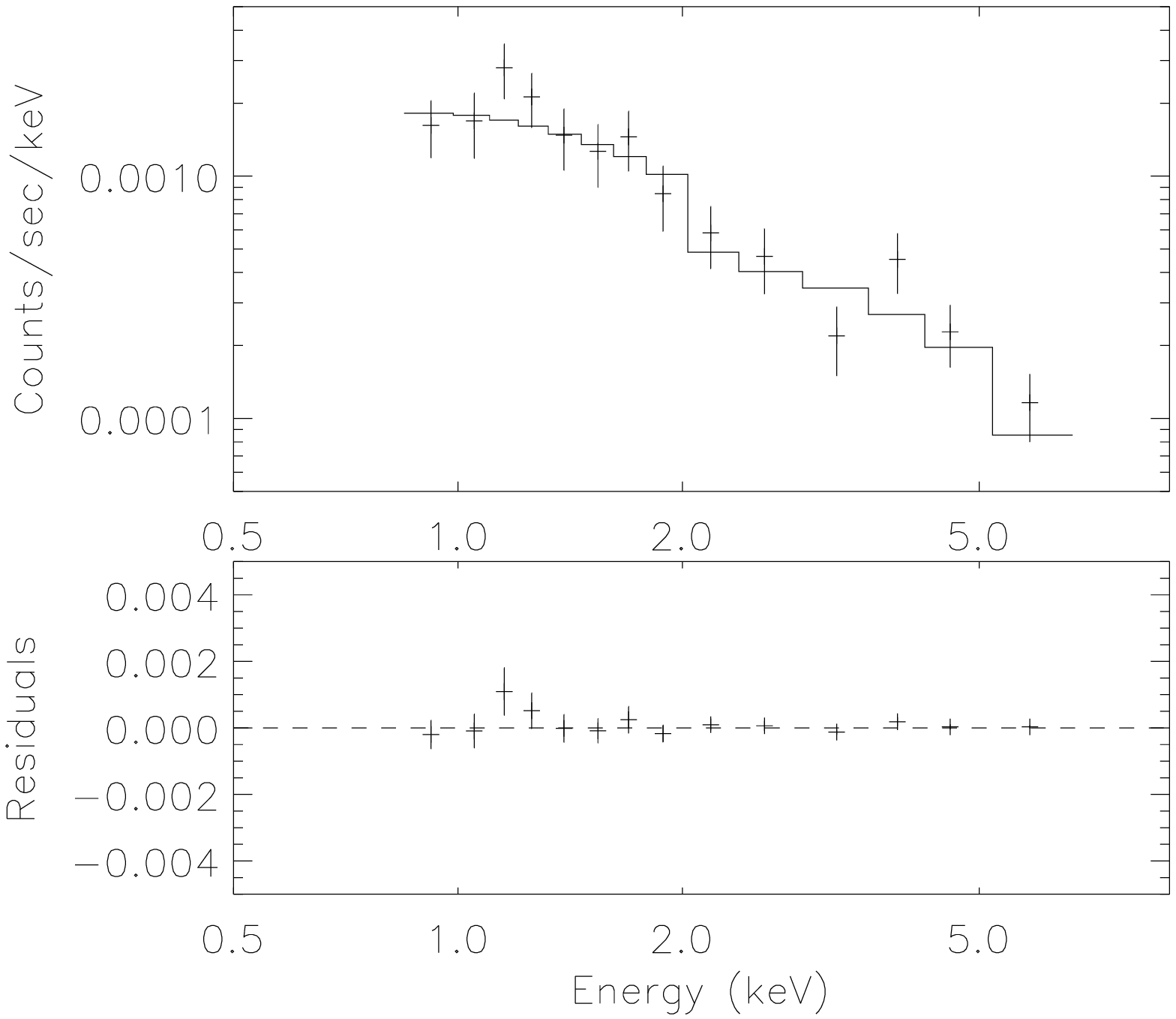}
\caption{The \chan\ spectrum of the core emission of \psr\ with best-fit 
blackbody plus power-law model (top) and magnetized, pure H atmospheric plus 
power-law model (bottom).  Also shown are the residuals from comparison of 
the data to the model in each case.}
\label{fig:chan_mod1}
\end{figure}

For the blackbody plus power-law model the best-fit parameters are a 
power-law index of $\Gamma=1.0^{+0.2}_{-1.0}$ and temperature of $T = 
(2.32^{+1.16}_{-0.81}) \times 10^{6}$ K.  Using a distance to the source 
of $D=1.04^{+0.21}_{-0.16}$ kpc \citep{chat04}, this implies a blackbody 
emitting radius of $0.12^{+0.16}_{-0.07}$ km.  This value is too small to be 
reconciled with the radius of the neutron star and would indicate that the 
origin of the emission is a hot polar cap.  We find an unabsorbed 0.3--10 keV
energy flux of $(6.4^{+19.3}_{-1.1}) \times 10^{-14}$ ergs cm$^{-2}$ s$^{-1}$.
The magnetized, pure H atmospheric plus power-law model best-fit parameters 
are a power-law index of $\Gamma=1.5^{+0.5}_{-0.4}$, temperature of 
$T^{\infty}_{eff} = (0.45^{+0.20}_{-0.22}) \times 10^{6}$ K, and a radius for 
the neutron star of $R_{NS}=7.2^{+7.2}_{-2.2}$ km.  For this fit the distance 
to the source and the mass of the neutron star were fixed at $D=1.04$ kpc and 
$M_{NS} = 1.4 M_{\sun}$, respectively.  The magnetic field of the neutron star
was fixed at $B=10^{12}$ G (this is a good approximation since the pulsar 
magnetic field as inferred from radio timing properties is $B = 8.4 \times 
10^{11}$ G, \citealt{hob04,man05}).  The unabsorbed 0.3--10 keV energy flux
for this fit is $(1.5^{+54.0}_{-0.7}) \times 10^{-13}$ ergs cm$^{-2}$ s$^{-1}$.
The \chan\ spectrum of the core emission from \psr\ is shown in 
Figure~\ref{fig:chan_mod1} with the best-fitting blackbody plus power-law model
(top) and magnetized, pure H atmospheric plus power-law model (bottom).

To analyze the compact diffuse emission we created a new events file in which 
the emission from the pulsar core was removed.  We extracted a spectrum for 
the diffuse component from a rectangular region of $40\arcsec \times 
55\arcsec$, centered on the emission and orientated along the direction of 
the pulsar's proper motion.  The background was extracted from a region of 
similar size offset from the diffuse emission.  The diffuse component 
extraction region contains 1207 counts, with an estimated 414 counts due to 
background.  We created the RMF and ARF files using standard CIAO tools.  
Before fitting the spectrum we regrouped the data, requiring a minimum of 15 
counts per spectral bin.  We modeled the spectrum over $0.5-7.0$ keV with an 
absorbed power-law, keeping the column density fixed at $N_{H}=0.50 \times 
10^{22}$ cm$^{-2}$.  The best-fit has a power-law index of $\Gamma=1.4\pm0.3$
and unabsorbed 0.3--10 keV energy flux of $(1.7^{+0.8}_{-0.5}) \times 10^{-13}$
ergs cm$^{-2}$ s$^{-1}$.

\begin{figure}
\plotone{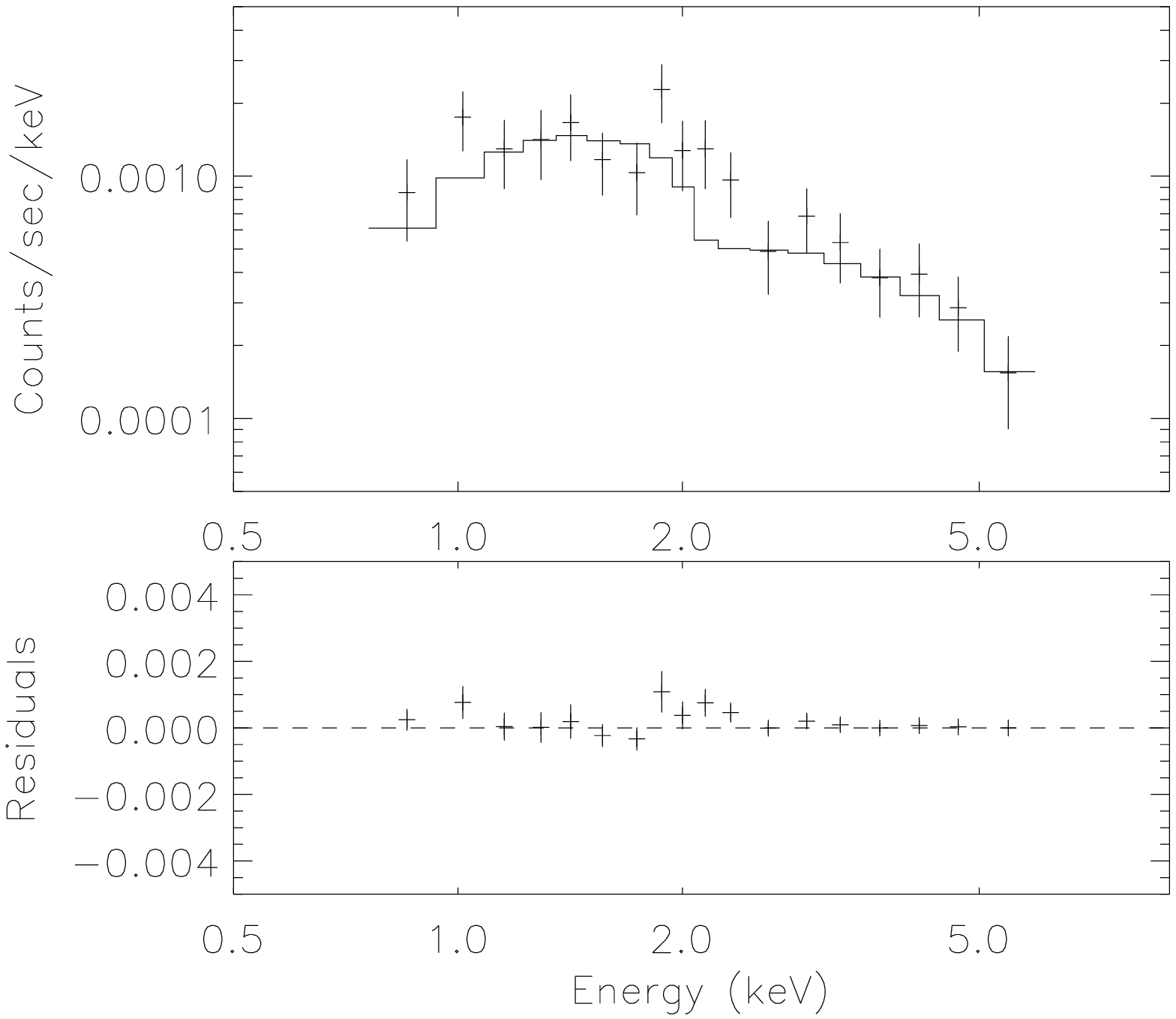}
\plotone{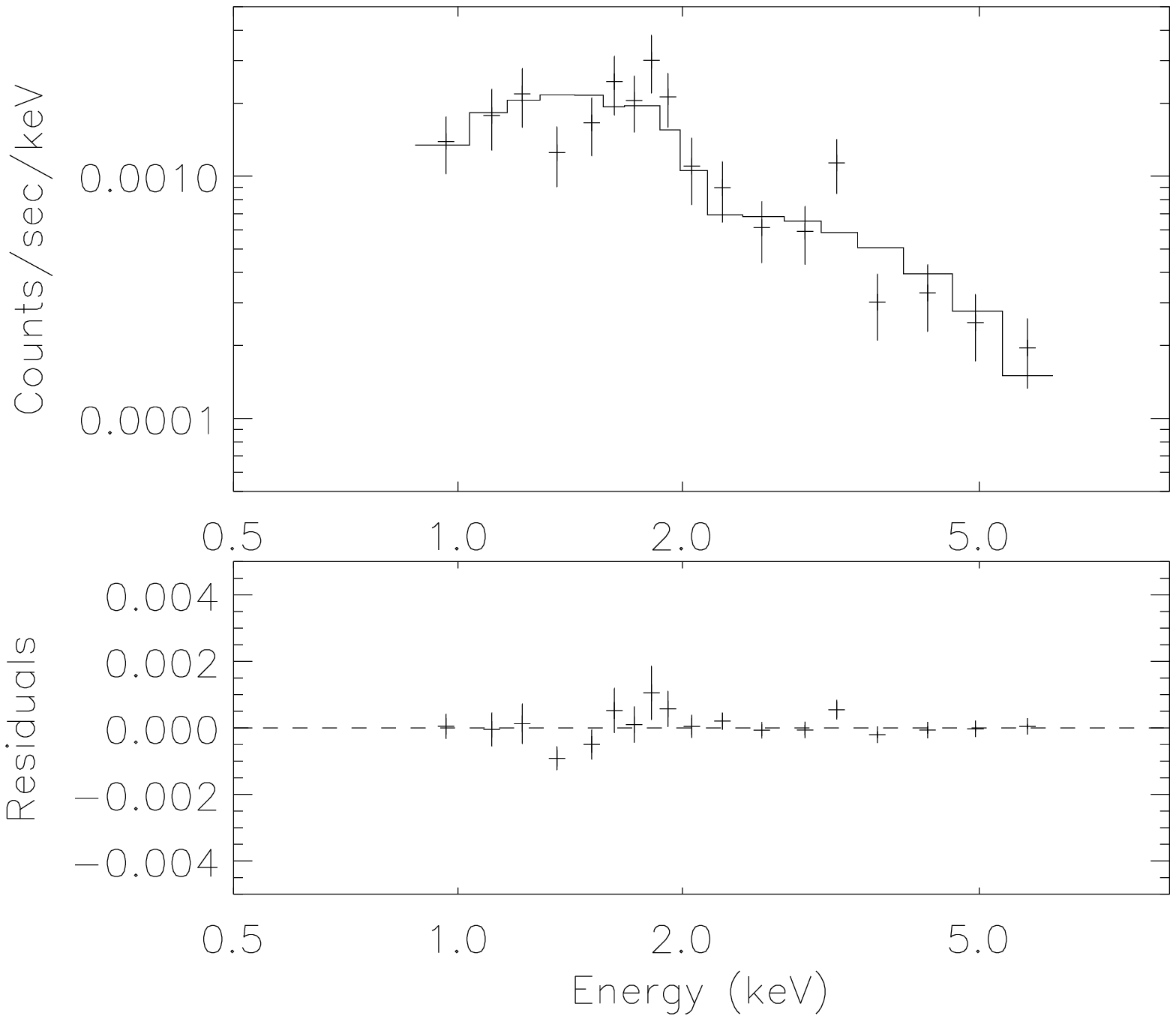}
\plotone{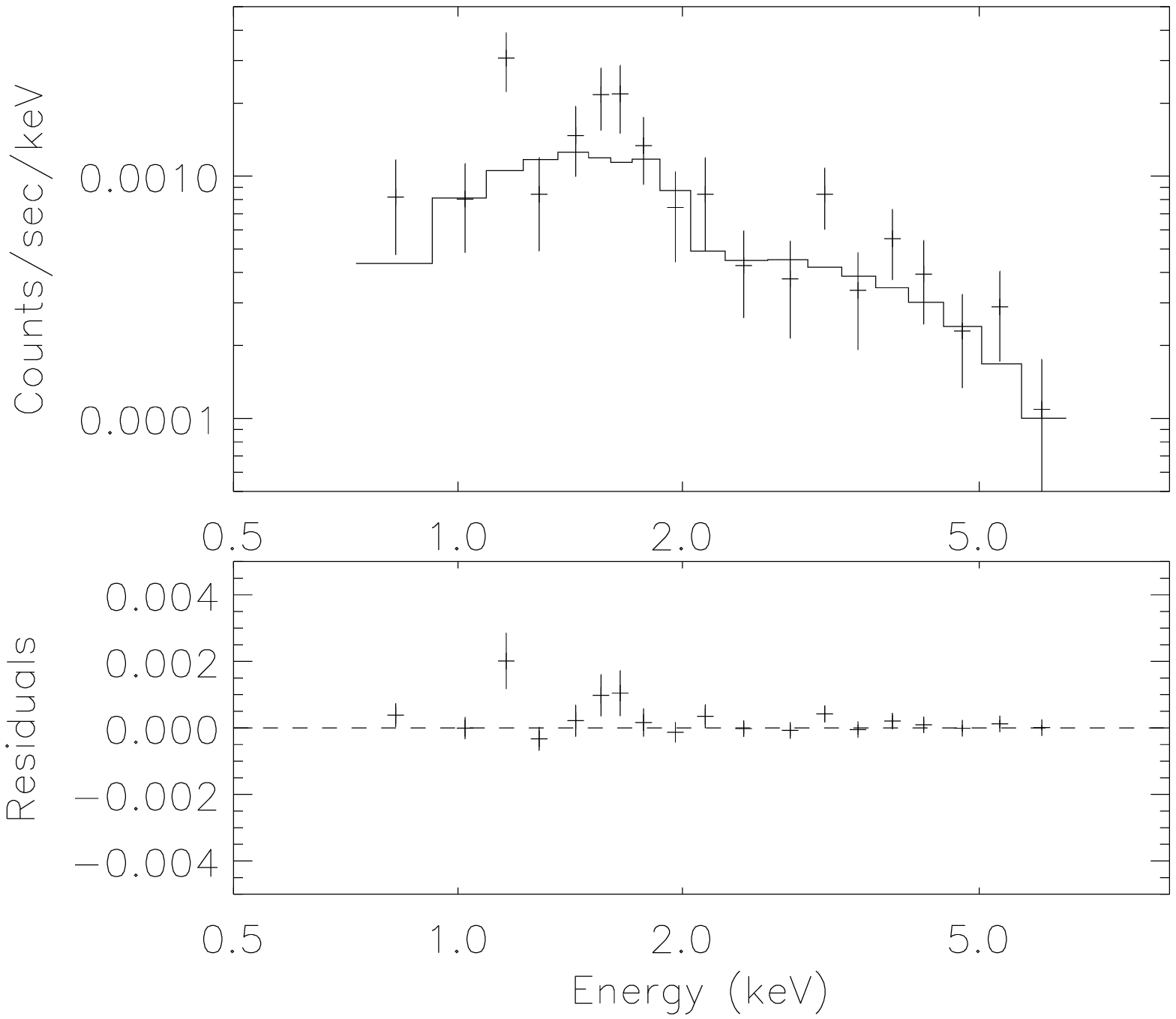}
\caption{Spectral fitting to \chan\ data of \psr.  First--third panels: 
diffuse emission from regions 1--3, respectively, with best fit power-law 
models.  Also shown are the residuals from comparison of the data to the 
model in each case.}
\label{fig:chan_mod2}
\end{figure}

In order to investigate the possibility of spectral evolution along the 
extended X-ray emission we created spectra for three regions of the compact 
diffuse emission.  The sizes of the regions were chosen with the aim of having 
similar numbers of counts in each region.  The three extraction regions, 
orientated along the direction of proper motion, are as follows, region 1: 
$40\arcsec \times 18\arcsec$, contains a total of 396 counts with 130 
attributed to background, region 2: $40\arcsec \times 9\arcsec$, contains a 
total of 356 counts with 69 attributed to background, region 3: $40\arcsec 
\times 28\arcsec$, contains a total of 454 counts with 214 attributed to 
background (see Figure \ref{fig_im} (middle panel)).  The background was 
extracted from the same region as above.  We created response files for each 
region and regrouped the spectra, requiring a minimum of 15 counts per 
spectral bin.  The spectra were fitted in the $0.5-7.0$ keV energy range with 
a power-law and fixed column density of $N_{H}=0.50 \times 10^{22}$ cm$^{-2}$. 
We find the best-fit power-law indices for regions 1--3 are 
$\Gamma=1.4\pm0.4$, $1.5\pm0.3$ and $1.2^{+0.5}_{-0.4}$, respectively.  Due to 
the uncertainties on the indices the presence of any spectral variability 
remains unclear.  The unabsorbed 0.3--10 keV energy fluxes are 
$(5.7^{+3.3}_{-1.8}) \times 10^{-14}$, $(7.6^{+3.4}_{-1.9}) \times 10^{-14}$ 
and $(5.3^{+4.4}_{-2.0}) \times 10^{-14}$ ergs cm$^{-2}$ s$^{-1}$, 
respectively.  The \chan\ spectra of the diffuse emission from regions 1--3 
are shown in Figure~\ref{fig:chan_mod2} (first--third panel) with the 
best-fitting power-law models.

We have also tried to determine if there are any spectral changes by using 
the hardness ratio $h_{2.0}$, which is defined as the ratio of counts above 
2.0 keV to that below 2.0 keV.  For the whole of the compact diffuse component
we find $h_{2.0} = 0.97 \pm 0.09$.  Regions 1, 2 and 3 have $h_{2.0} = 1.05 
\pm 0.17$, $0.91 \pm 0.13$ and $0.97 \pm 0.18$, respectively.  Due to the 
uncertainties, no particular trends in hardness ratio can be determined from
one region to the next.

\section{Timing Analysis}
\label{timing}

We barycentrically corrected the photon arrival times in the \xmm\ PN event 
file before performing the temporal analysis.  We extracted data for the source
from circular regions of $15\arcsec$ and $30\arcsec$ centered on the pulsar
position.  The total counts encompassed in these regions are 391 and 1143 
respectively, with the background contributing 151 and 562 counts, 
respectively.

In order to search for an X-ray modulation at the \psr\ spin frequency, we 
first determined a predicted pulse frequency at the epoch of our \xmm\ 
observations, assuming a linear spin-down rate and using the radio 
measurements \citep{hob04,man05}. We calculate $f = 6.3945388$ Hz at the 
midpoint of our observation (MJD 52315.7).  As glitches and/or deviations 
from a linear spin-down may alter the period evolution, we then searched for 
a pulsed signal over a wider frequency range centered on $f = 6.39454$ Hz.  We 
searched for pulsed emission using two methods.  In the first method we 
implement the $Z^{2}_{n}$ test \citep{buc83}, with the number of harmonics 
$n$ being varied from 1 to 5.  In the second method we calculate the Rayleigh 
statistic \citep{dej91,mar72} and then calculate the maximum likelihood 
periodogram (MLP; see e.g. \citealt{za02}) using the $C$ statistic 
\citep{cas79} to determine significant periodicities in the data sets.

The frequency search of the data extracted from an aperture of radius 
$30\arcsec$ does not yield any significant peaks near to the predicted 
frequency with either search method.  From our spatial analysis we know that 
the core of \psr's emission lies within $\sim 5\arcsec$ of the pulsar 
position.  We have therefore also searched for pulsed modulations in data 
extracted from a smaller aperture, however we have used a radius of 
$15 \arcsec$ as any smaller does not encompass enough counts for a meaningful 
analysis.

\begin{figure}
\plotone{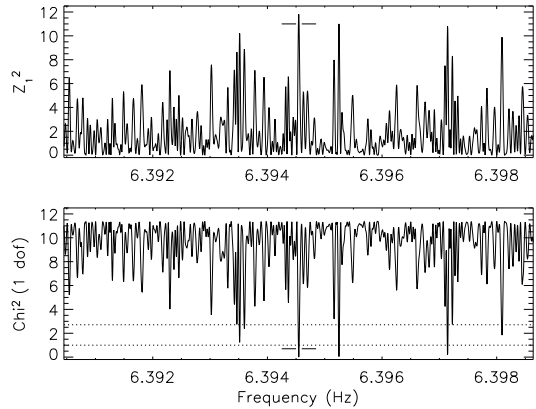}
\caption{$Z^{2}_{1}$-test (top) and Maximum Likelihood Periodogram (MLP; 
bottom) for the PN data of \psr.  The dominant peak from the $Z^{2}_{1}$-test 
and the corresponding peak in the MLP are marked.  The peaks occur at
$6.3945447^{+0.0000167}_{-0.0000107}$ Hz and 
$6.3945467^{+0.00000821}_{-0.00000818}$ Hz, respectively.  The dotted lines 
represent the 68\% ($\chi^{2}=2.71$) and 90\% ($\chi^{2}=1.0$) confidence 
levels for the frequencies in the MLP.}
\label{fig:psearch}
\end{figure}

The most significant $Z^{2}_{n}$-statistic occurs for $n=1$.  With the number 
of harmonics equal to one, the $Z^{2}_{n}$-statistic corresponds to the well 
known Rayleigh statistic.  We find three peaks with $> 90$\% significance in 
the MLP, all with corresponding peaks from the $Z^{2}_{1}$-test (see 
Figure~\ref{fig:psearch}).  The dominant peak from the $Z^{2}_{1}$-test occurs 
at $6.3945447^{+0.0000167}_{-0.0000107}$ Hz, with the corresponding peak in 
the MLP occurring at $6.3945467^{+0.00000821}_{-0.00000818}$ Hz.  The 
uncertainties quoted are the 68\% confidence limits on the position of the 
peak.  Both frequencies are consistent, within the 68\% contour, with the 
predicted pulse frequency, and with each other within the 90\% contour.  The
second most prominent peak from the $Z^{2}_{1}$-test, and the corresponding 
peak in the MLP, are not consistent with the predicted pulse frequency.  

\begin{figure}
\plotone{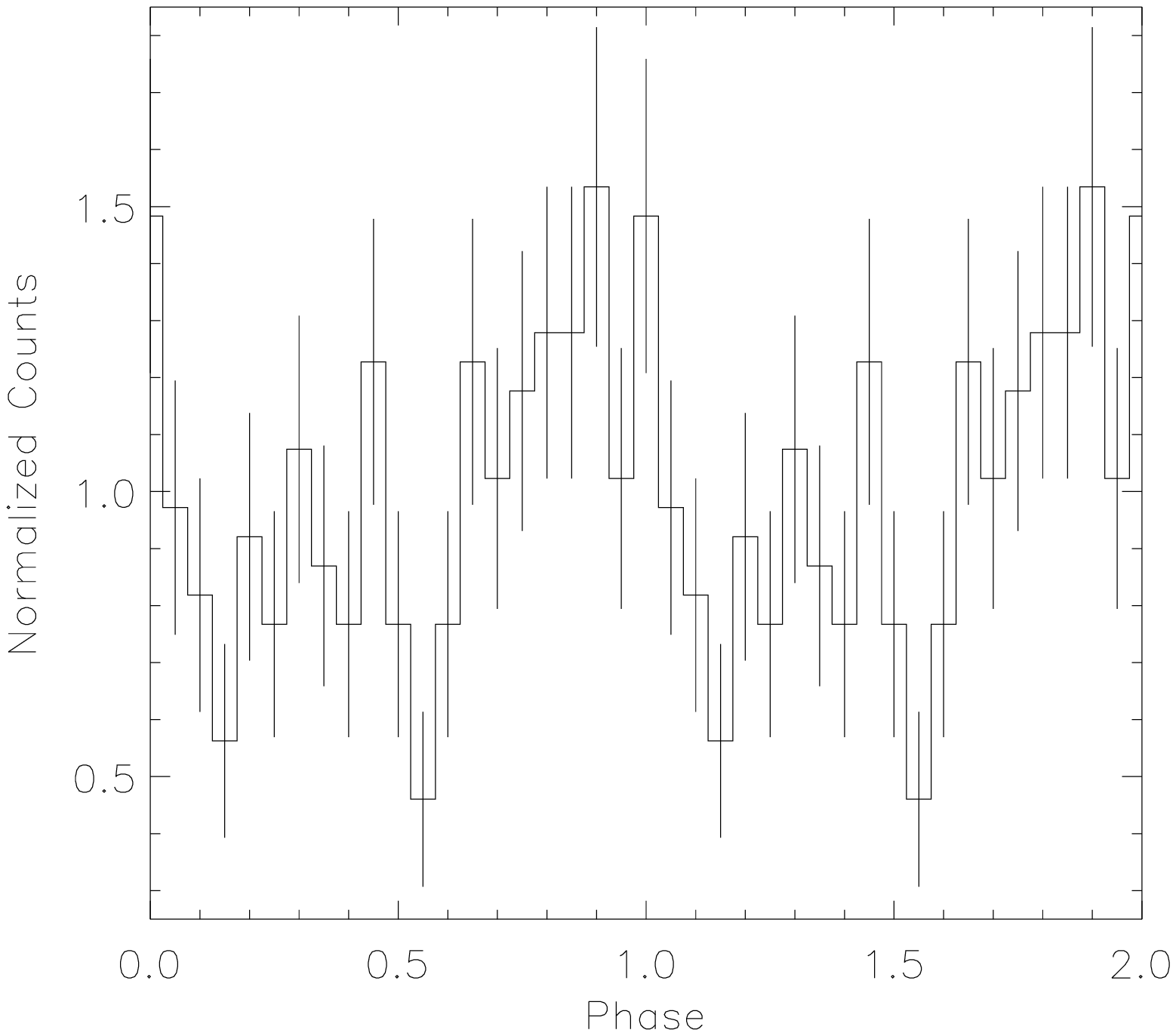}
\plotone{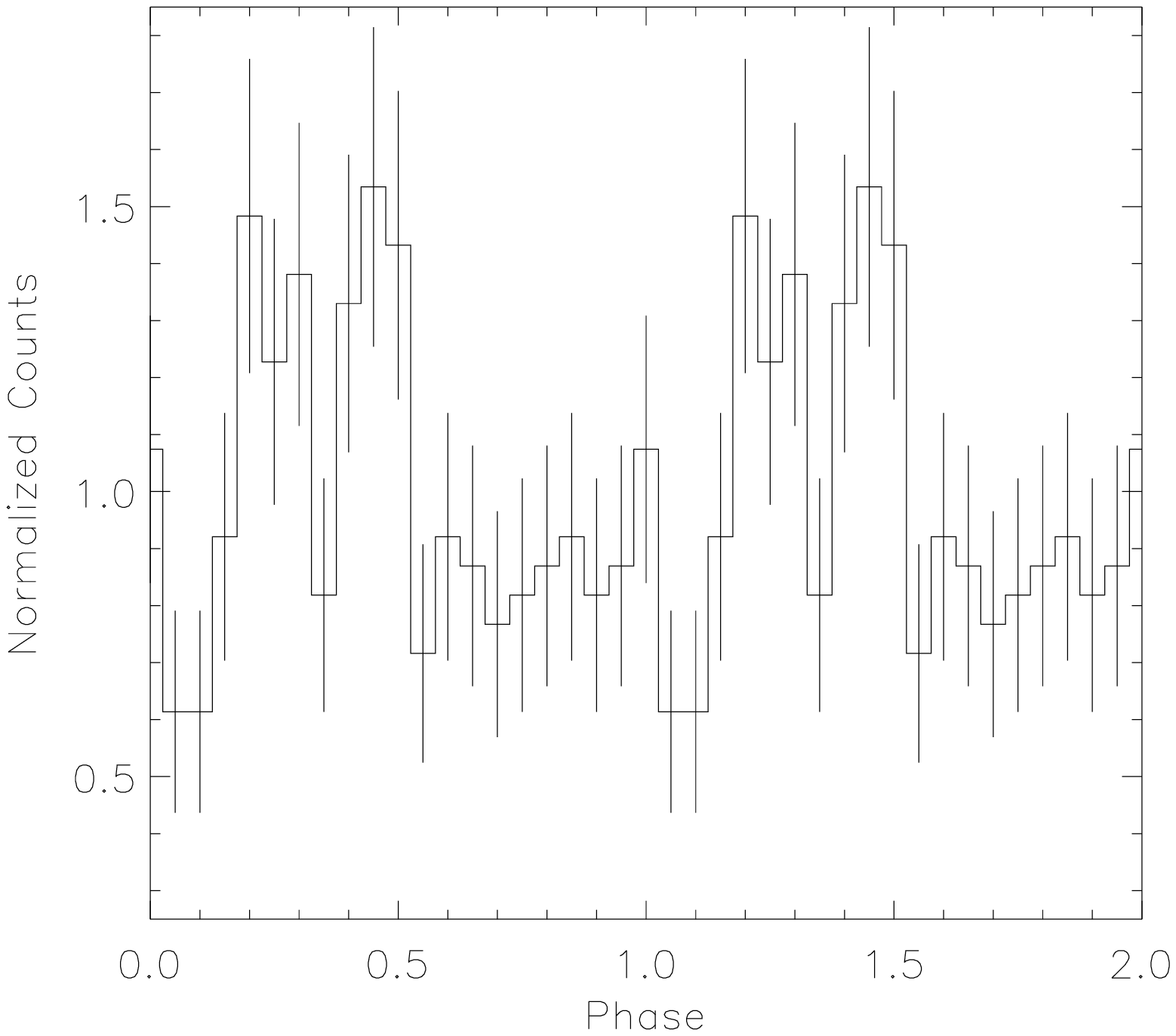}
\caption{PN data in the 0.3--10 keV energy range for \psr\ folded on the
frequency predicted from the radio measurements (top) and the frequency 
found from the $Z^{2}_{1}$-test (bottom).  In both cases the data are folded
using the radio ephemeris.}
\label{fig:fold}
\end{figure}

While we have detected a frequency that is in agreement with the predicted
pulse frequency for \psr\ we caution that the $Z^{2}_{1}$ peak has a 
probability of chance occurrence of $3\times 10^{-3}$.  Further observations 
of the source are needed to show whether the modulation detected is in fact 
pulsed X-ray emission from \psr.  We have folded the data on the predicted 
pulse frequency and the frequency found from the $Z^{2}$-test (see Figure 
\ref{fig:fold}); by fitting the profiles with a sinusoid we find that the 
modulation amplitude for the former is $25 \pm 7$\% and $21 \pm 8$\% for 
the latter.

\section{Discussion}
\label{disc}

Our spatial analysis of the \xmm\ and \chan\ observations of \psr\ have not 
only revealed X-rays from the pulsar, but have provided definitive proof of 
diffuse emission extending in the opposite direction to the pulsar's proper 
motion.  Similar detections of extended emission have been seen for other 
sources (e.g. N157B, \citealt{wan98}; PSR B1757-24, \citealt{fra91,kas01}; PSR 
B1957+20, \citealt{stap03}; PSR B1951+32, \citealt{li05}), and have been 
interpreted as emission from a ram-pressure confined PWN.  

We cannot separate the core and diffuse emission components for the \xmm\ data
and find that the spectrum can be well-fitted with a power-law model with
index $\Gamma=1.5^{+0.5}_{-0.3}$, similar to the value found for other PWNe 
\citep{kas05}.  The nebular emission is most likely dominating the spectrum.
The core emission from the \chan\ data can be well-fitted by a thermal plus
power-law model.  A fit with a blackbody plus power-law model gives 
$T=(2.32^{+1.16}_{-0.81}) \times 10^{6}$ K and $\Gamma =1.0^{+0.2}_{-1.0}$.
The fitted blackbody flux corresponds to an emitting radius of 
$0.12^{+0.16}_{-0.07}$ km, where the distance to the source is 1.04 kpc.  In 
this case, the size of the emitting region implies that the flux originates 
from a hot polar cap.  We can also fit the spectrum with a pure H, magnetized
atmospheric plus power-law model.  With the distance to the source fixed
at 1.04 kpc, this fit results in $T_{eff}^{\infty} = (0.45^{+0.20}_{-0.22}) 
\times 10^{6}$ K, $\Gamma=1.5^{+0.5}_{-0.4}$ and $R_{NS}=7.2^{+7.2}_{-2.2}$ km.
Taking into account the possible detection of pulsed emission from \psr\ 
it is likely that the emitting region is a hot polar cap.

It is suggested that the presence of a PWN is related to the spin-down power 
of the pulsar, and for sources with $\log \dot{E} \lesssim 36$ the PWN 
emission efficiency is significantly reduced \citep{fra97,gae00,got03}.  For
\psr\ $\log \dot{E} = 34.6$, making it one of a handful of sources with 
spin-down power below this limit with a detectable PWN (cf. Geminga, 
\citealt{car03}).  Using the results from the blackbody plus power-law fit to 
the core emission detected with \chan\ we determine an isotropic unabsorbed 
luminosity in the 0.3--10 keV band of $8.3 \times 10^{30}$ erg s$^{-1}$.  With 
$\dot{E} = 4.5 \times 10^{34}$ erg s$^{-1}$ for \psr, this leads to a 
conversion efficiency of $2 \times 10^{-4}$.  So in fact, we find that the 
conversion efficiency of the point source is similar to the values found for 
other pulsars \citep[see e.g.][]{bec97,gae04}.  Our analysis also indicates
that the compact diffuse component is more luminous than the point source, 
with a conversion efficiency of $5 \times 10^{-4}$ in the 0.3--10 keV range.
This result is again consistent with other sources \citep{bec97}.  In addition,
it is reported that when the pulsar spin-down energy is $\log \dot{E} 
\lesssim 36.5$ the morphology of the PWN seems to transition from toroidal to 
a jet/tail \citep{kas05}.  Our measurements of \psr\ appear to agree with 
this trend.

The morphology of the diffuse emission depends on how the interaction with 
the interstellar medium (ISM) or supernova remnant constrains the flow of 
particles \citep[e.g.][]{rey84}.  For a pulsar that is moving with a 
supersonic space velocity, the interaction of the supersonic flow with the 
ambient medium causes the speed of the flow to decrease sharply, while the 
density increases, forming a bow shock.  In addition to the bow shock, which 
is at some distance ahead of the pulsar, a reverse shock is formed nearer to 
the source which terminates the pulsar wind.

The results of the spatial analysis of the \xmm\ PN and \chan\ ACIS data of 
\psr\ indicate that the bulk of the diffuse emission extends $\sim 50\arcsec$ 
[0.25($d/1.04$ kpc) pc] downstream from the pulsar.  Using the measurements 
of the pulsar's proper motion \citep{chat04} we find that the transverse 
velocity of \psr\ is $v_{t} = 61$ km s$^{-1}$.  This implies that the time 
taken for the pulsar to have traversed the length of the diffuse emission is 
$>4000$ yr.  In addition, by considering the analysis of the \xmm\ MOS1 data 
we find that the diffuse emission may extend as far as $\sim 5\arcmin$ 
[1.51($d/1.04$ kpc) pc] from the point source.  This results in a travel time 
of $>24000$ yr for the pulsar.  Following the work of \citet{wan98} (see also 
\citealt{kas01}) the synchrotron lifetime of an electron of energy $E$ (in 
keV) can be defined as $t_{s} \sim 40 E^{-1/2}B_{-4}^{-3/2}$ yr, where 
$B_{-4}^{-3/2}$ is the magnetic field in units of $10^{-4}$ G.  Assuming that 
the dominant loss mechanism is synchrotron emission, i.e. $B>3.2$ $\mu$G, and 
that the energy of the photon is $E\sim 5$ keV, then $t_{s} \sim 3000$ yr.  
This indicates that the diffuse emission that we detect is not due to 
particles deposited by the pulsar as it traveled through space.  Hence, there 
must be a constant supply of wind particles traveling at velocities greater 
than the space velocity of the pulsar.  In addition, the particle flow 
velocity must be high enough such that the time for the flow to cross the 
length of the diffuse emission is less than the radiative lifetime of the 
particles.

Using the \chan\ data we have modeled the spectrum of the compact diffuse 
emission, excluding the contribution from the pulsar, finding that the data 
can be well-fitted with a power-law.  In other sources the power-law is seen 
to soften as one moves away from the pulsar position 
\citep[see e.g.][]{sla02,li05,kas05}.  An increase in the spectral index is 
expected as the particles will be cooler, i.e. older, at greater distance from 
the pulsar.  Our results indicate that we are detecting relatively hard 
emission, but due to the uncertainties, we are unable to comment on any 
changes in the spectral slope.  To measure cooling the PWN must be of an 
adequate size, it may be that for \psr\ the compact diffuse region is not 
large enough for a substantial change in power-law index to be measured, and
there are too few counts in the more extended diffuse region to perform a 
spectral analysis.  It is noted however that by comparing the spectral indices 
from the blackbody plus power-law fit to the core emission and the power-law 
fit to the compact diffuse emission we do detect an increase in $\Gamma$ of 
$\sim 0.5$.

\citet{gae04} have presented a detailed analysis of the diffuse X-ray emission 
associated with the radio source G359.23-0.82, also known as ``the Mouse''.
Their hydrodynamic simulations show that there are a number of regions that 
can be defined in a pulsar bow shock.  These include a pulsar wind cavity, 
shocked pulsar wind material, contact discontinuity (CD) and shocked ISM.  

The energetic shocked particles from the pulsar are confined by the CD, the 
position of which denotes the transition to the shocked ISM.  Following the 
method of \citet{gae04} we have estimated the distance between the peak of the
emission from \psr\ and the sharp cut-off in brightness ahead of the pulsar.  
Using the same limit as \citet{gae04} i.e. where the X-ray surface brightness 
falls by $1/e^{2} = 0.14$, we find a distance of $0.9\arcsec \pm 0.2 \arcsec$, 
giving the CD a projected radius of $r_{\rm CD} = 0.004 \pm 0.001$ pc.  Here, 
and in the following, we have used a distance to the pulsar of 1.04 kpc 
\citep{chat04}.  From Eq.~(1) of \citet{gae04} we can estimate the radius 
of the forward termination shock (TS), $r_{\rm TS}^{F} \sim 0.003$ pc.  This
corresponds to an angular distance of $\theta = 0.59 \arcsec$.  Comparing
our values to those for the Mouse implies that the emission in front of the 
pulsar is more compact in \psr\ than for the Mouse.  In both cases the close 
proximity of the forward TS to the peak X-ray emission renders the TS 
undetectable.  Using our results and Eq.~(2) of \citet{gae04} we find that 
\psr\ produces a ram pressure of $\rho v_{t}^{2} \sim 1.4 \times 10^{-9}$ 
ergs cm$^{-3}$.   Assuming cosmic abundances, this gives $v_{t} \sim 247 
n_{0}^{-1/2}$ km s$^{-1}$, where $n_{0}$ is the number density of the ambient 
medium, so for \psr\ we determine $n_0 \approx 0.06$ cm$^{-3}$, which is not 
unrealistic.  

Additional information can be obtained by equating the pressure of the pulsar 
wind (assumed isotropic), to that of the ambient medium.  By introducing the 
Mach number $M = v_{t}/ c_{s}$, where $c_{s}$ is the adiabatic sound speed in 
the ambient medium, and using the same prescription as \citet{gae04} for a 
representative ISM pressure (i.e. $P_{ISM}= 2400 k P_0$ erg cm$^{-3}$, with 
$0.5 \lesssim P_{0} \lesssim 5$ and $k$ is the Boltzmann's constant), this 
gives: 
\begin{equation} 
\dot E/[4 \pi (r_{TS}^{F})^{2} c] = 2400 k \gamma_{ISM} P_0 M^2 \, , 
\end{equation} 
from which we can obtain an estimate of the Mach number.  We find that for
\psr\ the sound speed $c_{s}$ of the medium lies in the range $1 - 30$ km 
s$^{-1}$.  The three principal phases of the ISM are generally named cold, 
warm and hot and are characterized by typical sound speed values of 1, 10 or 
100 km s$^{-1}$; according to this denomination our result implies that the 
pulsar is moving in either a cold or mildly warm ambient gas.  For comparison, 
in the case of the Mouse, \citet{gae04} found that the most probable pulsar 
velocity requires that the pulsar is moving through a warm phase of the ISM.  

\citet{gae04} also discuss the possible detection of the backward TS in their 
data.  Their simulations show that this feature has a closed surface, while 
the CD and bow shock are unrestricted.  The backward TS should lie much 
further away from the pulsar than the forward TS, i.e. $r_{\rm TS}^{B} \gg 
r_{\rm TS}^{F}$.  In principle this means that the backward TS may be 
detectable.  The possible dip we see in the profiles for the \psr\ data could 
indicate the presence of the backward TS.  The angular separation of the dip 
in our data is $\sim 10\arcsec$, a value consistent with that for the 
Mouse.  For a backward TS, \citet{gae04} predict that there would be a lack
of spectral evolution, a result we have found for the diffuse emission of
\psr.  However, we note that the feature in the Mouse data (and simulations) 
is quite compact in the north-south direction in comparison to the \psr\ 
feature.  In addition, the number of counts we detect for \psr\ may hinder
our investigation of the presence of such a feature.  Deeper observations
are needed to probe further the PWN of \psr.

A few days before submission of this paper, Tepedelenlio\v{g}lu \& \"Ogelman 
submitted a paper based on the same (public) observations to ApJL 
(astro-ph/0512209).

\acknowledgments

The authors wish to thank the referee for useful comments that have helped 
improve the paper.  This work is based on observations obtained with \xmm, 
an ESA science mission with instruments and contributions directly funded
by ESA Member States and NASA.  Support for this work was provided by 
the National Aeronautics and Space Administration through Chandra Award 
Number NNG04EF62I issued by the Chandra X-ray Observatory Center, which is 
operated by the Smithsonian Astrophysical Observatory for and on behalf of 
the National Aeronautics Space Administration under contract NAS8-03060.  
SZ thanks PPARC for its support through a PPARC Advanced Fellowship


\begin{thebibliography}{}

\bibitem[\protect\citeauthoryear{Becker \& Tr\"umper}{1997}]{bec97}
Becker, W., Tr\"umper, J. 1997, \aap, 326, 682
\bibitem[\protect\citeauthoryear{Bhat et al.}{1990}]{bha90}
Bhat, P. N., Acharya, B. S., Gandhi, V. N., Ramana Murthy, P. V., 
Sathyanarayana, G. P., Vishwanath, P. R. 1990, \aap, 236, 1
\bibitem[\protect\citeauthoryear{Buccheri et al.}{1983}]{buc83} 
Buccheri, R., et al. 1983, \aap, 128, 245
\bibitem[\protect\citeauthoryear{Caraveo et al.}{2003}]{car03}
Caraveo, P. A., Bignami, G. F., DeLuca, A., Mereghetti, S., Pellizzoni, A., 
Mignani, R., Tur, A., Becker, W. 2003, Science, 301, 1345
\bibitem[\protect\citeauthoryear{Cash}{1979}]{cas79} 
Cash, W. 1979, \apj, 228, 939
\bibitem[\protect\citeauthoryear{Chatterjee et al.}{2004}]{chat04}
Chatterjee, S., Cordes, J. M., Vlemmings, W. H. T., Arzoumanian, Z., 
Goss, W. M., Lazio, T. J. W. 2004, \apj, 604, 339
\bibitem[\protect\citeauthoryear{de Jager}{1991}]{dej91} 
de Jager, O. C. 1991, \apj, 378, 286
\bibitem[\protect\citeauthoryear{Frail \& Kulkarni}{1991}]{fra91}
Frail, D. A., Kulkarni, S. R. 1991, \nat, 352, 785
\bibitem[\protect\citeauthoryear{Frail \& Scharringhausen}{1997}]{fra97}
Frail, D. A., Scharringhausen, B. R. 1997, \apj, 480, 364
\bibitem[\protect\citeauthoryear{Gaensler et al.}{2000}]{gae00}
Gaensler, B. M., Stappers, B. W., Frail, D. A., Moffett, D. A., Johnston, S., 
Chatterjee, S. 2000, \mnras, 318, 58
\bibitem[\protect\citeauthoryear{Gaensler}{2001}]{gae01}
Gaensler, B. M. 2001, in Young Supernova Remnants, eds. S. S. Holt \& U. Hwang,
AIP Conference Proceedings, 565, 295
\bibitem[\protect\citeauthoryear{Gaensler et al.}{2004}]{gae04}
Gaensler, B. M., van der Swaluw, E., Camilo, F., Kaspi, V. M., Baganoff, 
F. K., Yusef-Zadeh, F., Manchester, R. N. 2004, \apj, 616, 383
\bibitem[\protect\citeauthoryear{Gotthelf}{2003}]{got03}
Gotthelf, E. V. 2003, \apj, 591, 361
\bibitem[\protect\citeauthoryear{Helfand}{1983}]{hel83}
Helfand, D. J. 1983, in IAU Symp. 101, Supernova Remnants and Their X-ray 
Emission, eds. J. Danzinger \& P. Gorenstein, Dordrecht: Reidel, 471
\bibitem[\protect\citeauthoryear{Hobbs et al.}{2004}]{hob04}
Hobbs, G., Lyne, A. G., Kramer, M., Martin, C. E., Jordan, C. 2004, \mnras,
353, 1311
\bibitem[\protect\citeauthoryear{Kaspi et al.}{2001}]{kas01}
Kaspi, V. M., Gotthelf, E. V., Gaensler, B. M., Lyutikov, M. 2001, \apj, 562,
163
\bibitem[\protect\citeauthoryear{Kaspi et al.}{2005}]{kas05}
Kaspi, V. M., Roberts, M. S. E., Harding, A. K. 2005, in Compact Stellar X-ray 
Sources, ed. W. H. G. Lewin \& M. van der Klis (Cambridge: Cambridge Univ. 
Press), in press
\bibitem[\protect\citeauthoryear{Li et al.}{2005}]{li05}
Li, X. H., Lu, F. J., Li, T. P. 2005, \apj, 628, 931
\bibitem[\protect\citeauthoryear{Manchester et al.}{2005}]{man05}
Manchester, R. N., Hobbs, G. B., Teoh, A., Hobbs, M. 2005, \aj, 129, 1993
\bibitem[\protect\citeauthoryear{Mardia}{1972}]{mar72} 
Mardia, K. V. 1972, Statistics of Directional Data (London: Academic)
\bibitem[\protect\citeauthoryear{Pavlov et al.}{1995}]{pav95}
Pavlov, G. G., Shibanov, Y. A., Zavlin, V. E., Meyer, R. D. 1995, in The Lives 
of Neutron Stars eds. A. Alpar, U. Kiliz\'oglu \& J. van Paradijs, Kluwer
Academic Publishers, p. 71
\bibitem[\protect\citeauthoryear{Pavlov et al.}{2001}]{pav01}
Pavlov, G. G., Kargaltsev, O. Y., Sanwal, D., Garmire, G. P. 2001, ApJ, 554, 
189
\bibitem[\protect\citeauthoryear{Rees \& Gunn}{1974}]{ree74}
Rees, M. J., Gunn, J. E. 1974, \mnras, 167, 1
\bibitem[\protect\citeauthoryear{Reynolds \& Chevalier}{1984}]{rey84}
Reynolds, S. P., Chevalier, R. A., 1984, \apj, 278, 630
\bibitem[\protect\citeauthoryear{Seward \& Wang}{1988}]{sew88}
Seward, F. D., Wang, Z.-R. 1988, \apj, 332, 199
\bibitem[\protect\citeauthoryear{Slane}{1994}]{sla94}
Slane, P. 1994, \apj, 437, 458
\bibitem[\protect\citeauthoryear{Slane et al.}{2002}]{sla02}
Slane, P. O., Helfand, D. J., Murray, S. S. 2002, \apj, 571, 45
\bibitem[\protect\citeauthoryear{Stappers et al.}{2003}]{stap03}
Stappers, B. W., Gaensler, B. M., Kaspi, V. M., van der Klis, M., Lewin, 
W. H. G. 2003, Sci, 299, 1372
\bibitem[\protect\citeauthoryear{Str\"uder et al.}{2001}]{str01} 
Str\"uder, L., et al. 2001, \aap, 365, L18
\bibitem[\protect\citeauthoryear{Tepedelenlio\v{g}lu \& \"Ogelman}{2005}]{tep05}
Tepedelenlio\v{g}lu, E., \"Ogelman, H. 2005, ApJL, submitted (astro-ph/0512209)
\bibitem[\protect\citeauthoryear{Turner et al.}{2001}]{tur01}
Turner, M. J. L., et al. 2001, \aap, 365, 27
\bibitem[\protect\citeauthoryear{Wang \& Gotthelf}{1998}]{wan98}
Wang, Q. D., Gotthelf, E. V. 1998, \apj, 494, 623
\bibitem[\protect\citeauthoryear{Weisskopf et al.}{2000}]{wei00}
Weisskopf, M. C. 2000, ApJ, 536, 81
\bibitem[\protect\citeauthoryear{Zane et al.}{2002}]{za02} 
Zane, S., et al. 2002, \mnras, 334, 345

\end{thebibliography}
\end{document}